\documentstyle[twocolumn,aps,epsfig]{revtex}

\begin{document}

%%%%%%%%%%%%%%%%%%%%%%%%

%twocol
\twocolumn[\hsize\textwidth\columnwidth\hsize\csname @twocolumnfalse\endcsname

%%\preprint{
\hfill$\vcenter{\hbox{\bf IFUSP/P-1323/98} 
                \hbox{\bf IFT-P.083/98} 
             }$ 
%%}

\title{Hunting a Light $U(1)_B$ Gauge Boson \\
           Coupled to Baryon Number in Collider Experiments}

\author{M.\ Drees$^1$\thanks{Email: drees@ift.unesp.br}, O.\ J.\ P.\ 
           \'Eboli$^1$\thanks{Email: eboli@ift.unesp.br}, J.\ K.\ 
           Mizukoshi$^2$\thanks{Email: mizuka@fma.if.usp.br} \\[8pt] }

\address{\em $^1$ Instituto de F\' {\i}sica Te\'orica -- UNESP \\
    R. Pamplona 145, 01405-900 S\~ao Paulo, Brazil \\[8pt]
    $^2$ Instituto de F\'{\i}sica, Universidade de S\~ao Paulo, \\
    C.\ P.\ 66.318, 05389-970 S\~ao Paulo, Brazil. }

\vskip 8pt

%\date{November 13, 1998}

\maketitle

%\vspace{.2in}

\hfuzz=25pt

\vskip -12pt

\begin{abstract}

%\vskip -24pt
  
  We analyze several signals at HERA and the Tevatron of a light
  $U(1)_B$ gauge boson ($\gamma_B$) coupling to baryon number. We show
  that the study of the production of $b \bar{b}$ pairs at the
  (upgraded) Tevatron can exclude $\gamma_B$ with masses ($m_B$) in
  the range $40 \lesssim m_B \lesssim 300$ GeV for $\gamma_B$
  couplings ($\alpha_B$) greater than $2 \times 10^{-2}$ ($3 \times
  10^{-3}$). We also show that the HERA experiments cannot improve the
  present bounds on $\gamma_B$. Moreover, we demonstrate that the
  production at HERA and the Tevatron of di--jet events with large
  rapidity gaps between the jets cannot be explained by the existence
  of a light $\gamma_B$.

\end{abstract}

%(un)comment for single colum
\vskip2pc]
\narrowtext

%\newpage

%******************************************************************************
\section{Introduction}

Global symmetries of the standard model (SM), like baryon and lepton
numbers, can be broken by quantum gravity effects \cite{qg}. In order
to avoid an unacceptably large proton decay rate induced by this
symmetry breaking we can elevate the conservation of baryon number to
a local symmetry \cite{mu:1}. Moreover, there are string models that
also gauge baryon number to protect the proton from decaying through
higher-dimensional operators. In particular, models with low string
scale, $E_{\rm string} \sim {\cal O}$(TeV), generally exhibit baryon
number as a local symmetry \cite{st:low}.

In this work, we analyze the signals of a light gauge boson coupling
to baryon number at the HERA and Tevatron colliders. We assumed that
the $U(1)_B$ symmetry is spontaneously broken, giving a mass ($m_B$)
to $\gamma_B$, and that the mixing between $\gamma_B$ and the
electroweak gauge bosons is negligible \cite{mu:2}. In the absence of
mixing, the $\gamma_B$ boson always decays into quark pairs; its
signature is thus the presence of 2 jets with an invariant mass close
to $m_B$. A previous analysis \cite{mu:3} studied its effect on
$Z$-pole physics and constrained its coupling to be $\alpha_B \lesssim
0.2$ for masses $m_B \le m_Z$. Moreover, $\Upsilon(1S)$ decays are
modified by the $\gamma_B$ boson, leading to stronger constraints on
$\alpha_B$ for $m_B \lesssim 30$ GeV \cite{mu:3}.

In our analysis we concentrate on the decay $\gamma_B \to b \bar{b}$
in order to reduce the QCD backgrounds. Using a muon trigger, we show
that the study of the production of $b \bar{b}$ pairs at the
(upgraded) Tevatron can exclude $\gamma_B$ bosons with masses ($m_B$)
in the range $40 \lesssim m_B \lesssim 300$ GeV for couplings
$\alpha_B \gtrsim 2 \times 10^{-2}$ ($3 \times 10^{-3}$). A displaced
vertex trigger could increase the sensitivity at the upcoming Main
Injector (MI) run by another factor of three. We also use the
available search by the CDF collaboration for particles decaying into
$b \bar{b}$ pairs \cite{cdf:bb}. Since this search uses a jet trigger
which needs to be pre-scaled if the jet energies are not very large,
the resulting limit turns out to be weaker than the sensitivity limit
of current data based on a muon trigger. We also analyze $\gamma_B$
production at HERA, however, the potential bounds derived from this
study are weaker than the $Z$--pole or Tevatron ones due to the
limited integrated luminosity.

In principle, $\gamma_B$ boson exchange in the $t-$channel can give
rise to events presenting a large rapidity gap between two jets
\cite{mu:3}. Recently D\O\ studied the gap fraction as a function of
$E_T$ \cite{jill}. Our fit to this data in terms of $\gamma_B$
exchange gives $\gamma_B$ masses very close to zero and $\alpha_B
\simeq 0.06$. However, this region is already excluded by the analysis
of $\Upsilon(1S)$ decays \cite{mu:3}. Moreover, the HERA data on
rapidity gaps can only be explained for $m_B \simeq 0$ and very large
$\alpha_B$. Therefore, the present data on the production of rapidity
gaps cannot be explained by the existence of a $\gamma_B$ boson.

In our analyses we assumed that the interaction between quarks and $\gamma_B$
is described by the effective Lagrangian
\[
{\cal L}_{\text{eff}} = \frac{1}{3} \sqrt{4 \pi \alpha_B}~ \bar{q} 
\gamma^\mu q~ B_\mu   \;\; ,
\]
where we denote the quark fields by $q$ and the $\gamma_B$ field by $B_\mu$.
Neglecting the fermions masses, the $\gamma_B$ width is given by
\[
\Gamma_B = \frac{N}{9}~ \alpha_B~  m_B \;\; ,
\]
where $N$ is the number of quark pairs to which $\gamma_B$ can decay.

%******************************************************************************

\section{Direct Searches at Tevatron and HERA}

In hadronic collisions $\gamma_B$ bosons can be produced in the $s$-channel
via quark--antiquark annihilation, leading to two-jet events. In order to
reduce the large QCD backgrounds we focused on $\gamma_B$ decays into $b
\bar{b}$ pairs ($p \bar{p} \to q \bar{q} \to \gamma_B \to b \bar{b} $). The
main backgrounds to this process are QCD $b \bar{b}$ production, mistagged QCD
jets, and $Z$ production for $m_B$ close to $M_Z$. Since the two-jet event
rate is too large for the present data acquisition system to be able to
analyze all events, we need to choose a trigger to select a subset of the
data. In our analyses of the Run I potential for $\gamma_B$ searches, we
considered only events containing a muon (from $b$ decay) with $p_T^{\mu} >
7.5$ GeV and $ |\eta_\mu| < 0.9$ (2 for the upgraded Tevatron) \cite{cdf:zbb}.
There is one useful byproduct of this requirement: together with the
requirement that both $b$'s be vertex tagged, the presence of this muon
reduces the mistag background to a negligible level. Moreover, we also
demanded that
\begin{itemize}

\item the jets should have $p_T > 15$ GeV;

        \item the jets should be separated by $\Delta R > 0.4$.

\end{itemize}
With these requirements the $b$ tag efficiency is 0.24. In our
analyses we evaluated the scattering amplitudes of the signal and
backgrounds using the package Madgraph \cite{madg}, taking into
account the interference between $\gamma_B$ and gluon$/Z$ exchange diagrams.

\begin{figure}
\begin{center}
  \mbox{\epsfig{file=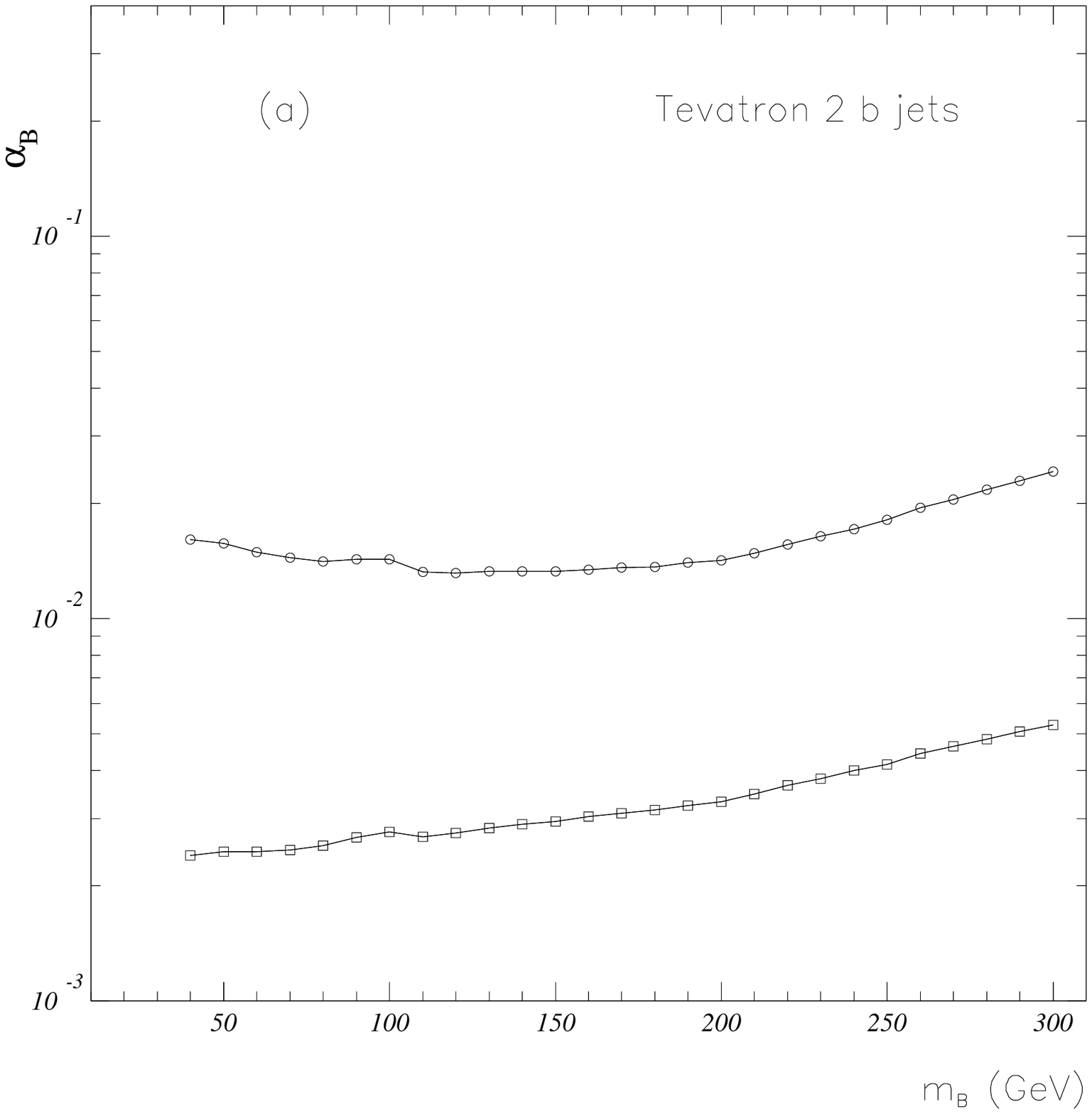,width=\linewidth}}
\end{center}                 
\begin{center}
  \mbox{\epsfig{file=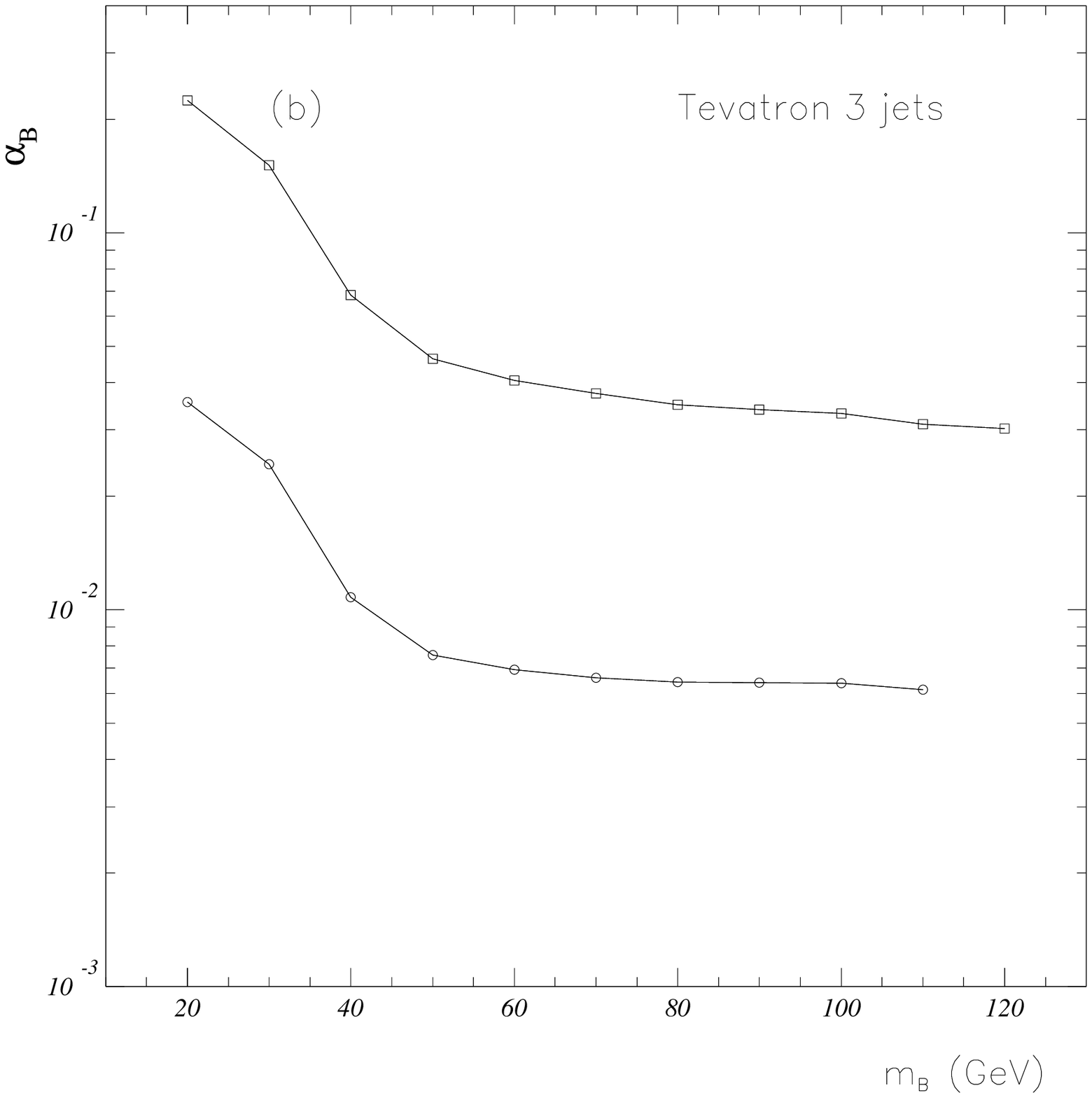,width=\linewidth}}
\end{center}
%%
%\begin{center}
%\parbox[c]{3.0in}{
%\mbox{\epsfig{file=fig1a.eps,width=\linewidth}}
%                 }
%\hfill
%\parbox[c]{3.0in}{
%\mbox{\epsfig{file=fig1b.eps,width=\linewidth}}
%                 }
%\end{center}
%%
\caption{The region that can be excluded at the $3\sigma$ level 
  using the Run I or Main Injector data is given above the upper and lower
  lines respectively. In (a) we considered $b \bar{b}$ final states while in
  (b) we studied $b \bar{b}+$jet events.}
\label{fig:bb}
\end{figure}

We considered that a point of the $(\alpha_B , m_B)$ plane is within
the reach of an experimental search, if the predicted signal has a
$3\sigma$ significance when we restrict ourselves to $b \bar{b}$
invariant masses in the range $ m_B - 10 \ {\rm GeV} < m_{b\bar{b}} <
m_B + 10$ GeV. We chose this rather high confidence level for an
exclusion limit since our analysis does not allow for experimental
resolutions or efficiencies (other than the $b-$tagging
efficiency). We note, however, that CDF recently found \cite{cdf:zbb}
preliminary evidence for $Z \to b \bar{b}$ decays in their Run I data
sample using cuts very similar to the ones applied by us. We show in
Fig.\ \ref{fig:bb}a the region in the $(\alpha_B , m_B)$ plane which
could be excluded by the CDF Run I data, {\it i.e.} for an integrated
luminosity of 110 pb$^{-1}$ and a $b$-tagging efficiency of 24\%, if
no signal is found. In this figure we also display the effect of
having a larger integrated luminosity (2 fb$^{-1}$) and an extended
rapidity acceptance for the muons ($|\eta_\mu | <2$) at the MI.
Notice that our $p_T$ and $\Delta R$ cuts constrain the invariant mass
of the $b \bar{b}$ pair in two-jet events to be larger than 30 GeV.

In order to explore smaller $\gamma_B$ masses we considered its production in
association with a jet. The processes that we analyzed are
\begin{eqnarray*}
p \bar{p} & \to \gamma_B g &\to b \bar{b} g \; ,
\\
p \bar{p} & \to \gamma_B q &\to b \bar{b} q \; ,
\end{eqnarray*}
where $q$ can be any quark or antiquark. We present in Fig.\
\ref{fig:bb}b the potential limits on $\alpha_B$ and $m_B$
originating from the study of $b$-$\bar{b}$-jet production for Run I
and at the MI. In this analysis we applied the same cuts for the $b
\bar{b}$ system and required the extra jet to have $p_T > 10 $ GeV and
to be separated from the $b$ jets by $\Delta R > 0.4$.

\begin{figure}
\begin{center}
\mbox{\epsfig{file=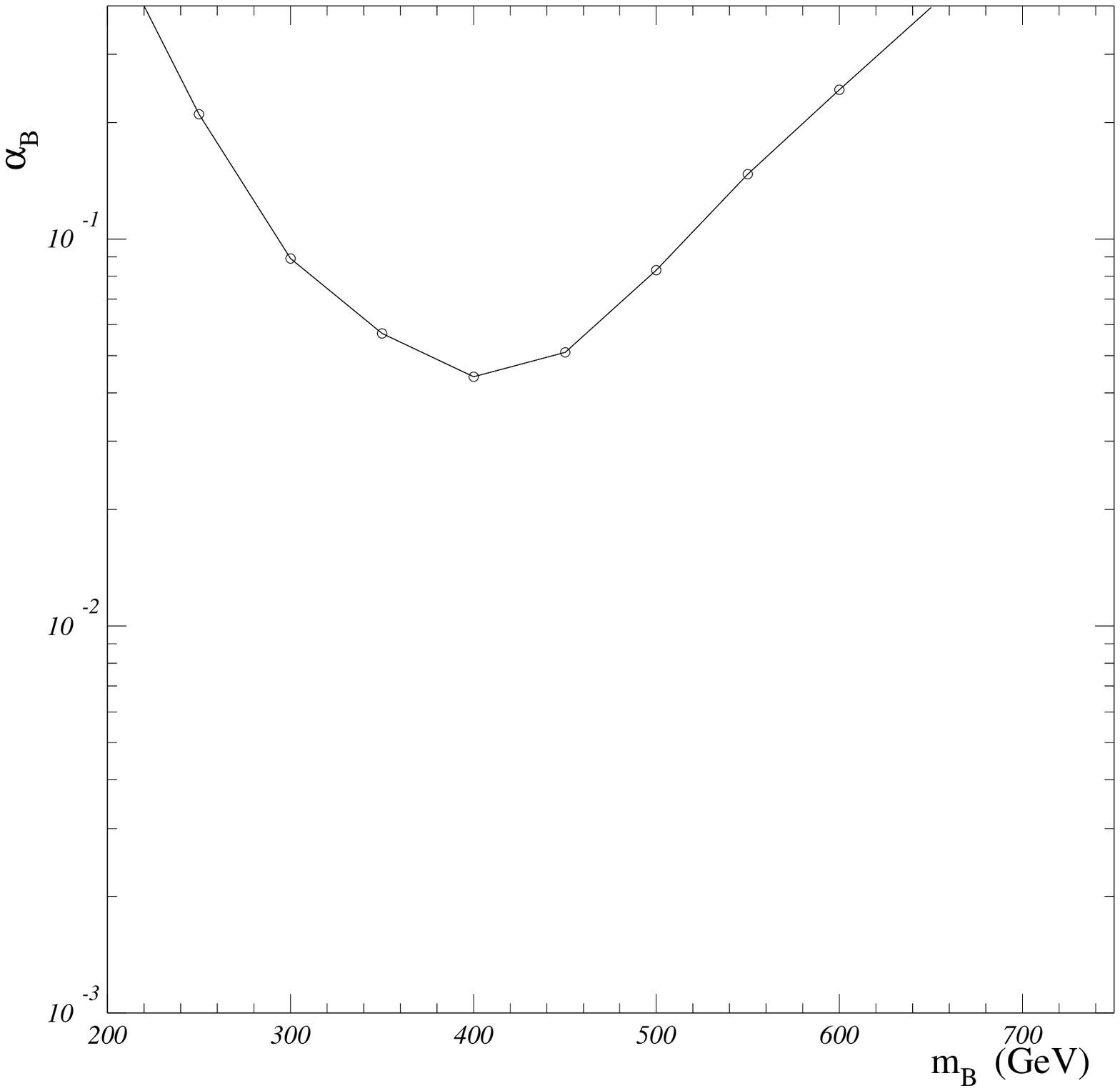,width=\linewidth}}
\end{center}
\caption{The region that can be excluded at 95\% CL from the CDF 
search \protect\cite{cdf:bb} using a jet trigger at Run I.}
\label{fig:cdf}
\end{figure}

It is also possible to search for $\gamma_B$ using the Run I data but
triggering on jets with a minimum $E_T$ \cite{cdf:bb}. This choice of
trigger requires pre-scaling, which leads to small effective
integrated luminosities at low values of $E_T$. Using the CDF excluded
production cross sections for particles decaying into $b \bar{b}$
pairs we obtained the limits on $\gamma_B$ shown in Fig.\
\ref{fig:cdf}. We emphasize that the bounds shown in this figure are
directly based on an experimental analysis, including all resolution
and efficiency effects. As we can see, for $m_B$ up to a few hundred
GeV the limits on $\gamma_B$ from this search are weaker than the ones
that should be obtainable using the muon trigger, if no signal is
found.

\begin{figure}
\begin{center}
\mbox{\epsfig{file=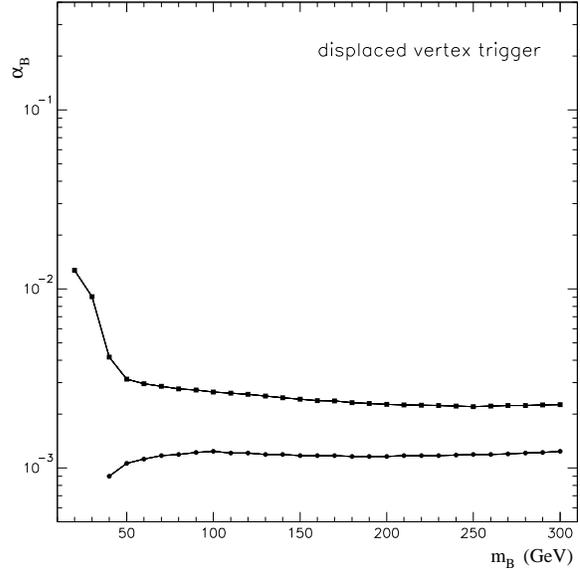,width=\linewidth}}
\end{center}
 \caption{The region that can be excluded at the $3\sigma$ level 
at the MI using a displaced vertex trigger. The upper (lower) line
is due to the $b$-$\bar{b}$-jet ($b\bar{b}$) production.}
\label{fig:dis}
\end{figure}

The Tevatron experimental collaborations are studying the possibility
of triggering events exhibiting displaced vertices for the upcoming
Main Injector run. We access the impact of this trigger on the
searches for $\gamma_B$ eliminating the cuts on the muon coming from
$b$ decays and introducing the QCD mistag background with a rate of
1\% per jet. All other cuts are unchanged. Fig.\ \ref{fig:dis}
contains the region of the $(\alpha_B , m_B)$ plane that can be probed
at the MI with this new trigger. As we can see from this figure, this
trigger can increase the sensitivity of the Tevatron for $\gamma_B$
searches.

%HERA results

At HERA, $\gamma_B$ bosons can be produced via the hadronic content of the
photon:
\[
  \gamma p \to q \bar{q} \to \gamma_B \to q^\prime \bar{q^\prime} \; .
\]
However, the two-jet signature of $\gamma_B$ is immersed in a large
background from resolved photons. It turns out that the signal can not
be observed even for the largest couplings allowed by the $Z$ physics
($\alpha_B \simeq 0.2$) for the presently available luminosities. In
order to observe a $\gamma_B$ signal for this large couplings one
would need an integrated luminosity of at least 250--500 pb$^{-1}$
depending on $m_B$. Therefore, the bounds on $\gamma_B$ from HERA are
much weaker than the $Z$ pole or Tevatron ones.

%******************************************************************************

\section{Rapidity gap analysis}

Experiments at HERA and the Tevatron have observed events containing
two jets with no hadronic activity between them. These occur with a
frequency of order of one percent in hadron--hadron collisions
\cite{cdf-prl}. This is just one example of an interaction mediated by
the exchange of the ``Pomeron'', a state which carries no net color.
Since $\gamma_B$ is a color singlet, it can also give rise to rapidity
gap events \cite{mu:3}. In this case the fraction of events presenting
rapidity gaps as well as their kinematical distributions are
determined by $\alpha_B$ and $m_B$, which allows us to constrain these
parameters. Here we extract the bounds on $\gamma_B$ from the study of
rapidity gaps assuming that these are only due to $\gamma_B$ exchange in
the $t-$channel.

The D\O\ Collaboration has recently measured the production cross
section of hard jets separated by a rapidity gap as a function of
transverse momentum and gap size \cite{jill}. This data indicates that
a large fraction of the gap events originates from quark--quark
collisions \cite{two}, a feature that is present in the $\gamma_B$
exchange. In order to obtain bounds on $\gamma_B$ from this data, we
evaluated
\[
F_{\text{gap}} (E_T) = \frac{ d \sigma_B / d E_T} 
{d \sigma_{\text{total}} / d  E_T} \; , 
\]
where $\sigma_B$ and $\sigma_{\text{total}}$ are the $\gamma_B$
contribution and the total cross section for the production of jet
pairs, respectively.

\begin{figure}
\begin{center}
 \mbox{\qquad\epsfig{file=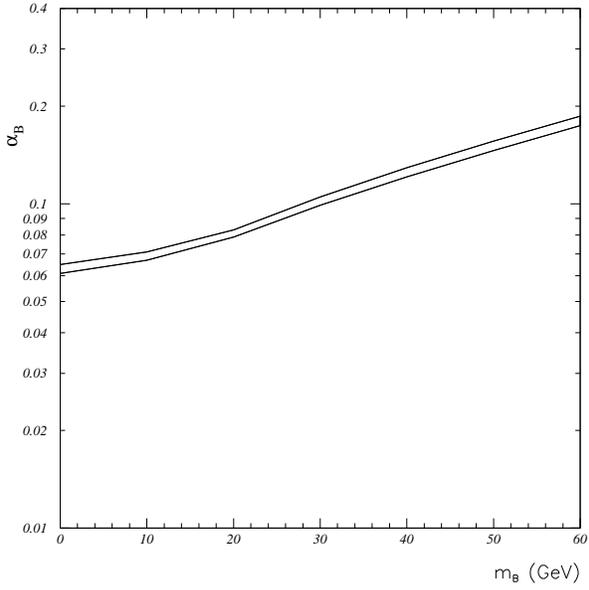,width=\linewidth}}
 \end{center}
 \caption{The region between the solid lines is the $1\sigma$ allowed 
   area obtained from our fitting procedure to the D\O\ data. We used the cuts
   and bins defined in Ref.\ \protect\cite{jill}.}
\label{tev:gap}
\end{figure}

\begin{figure}
\begin{center}
 \mbox{\qquad\epsfig{file=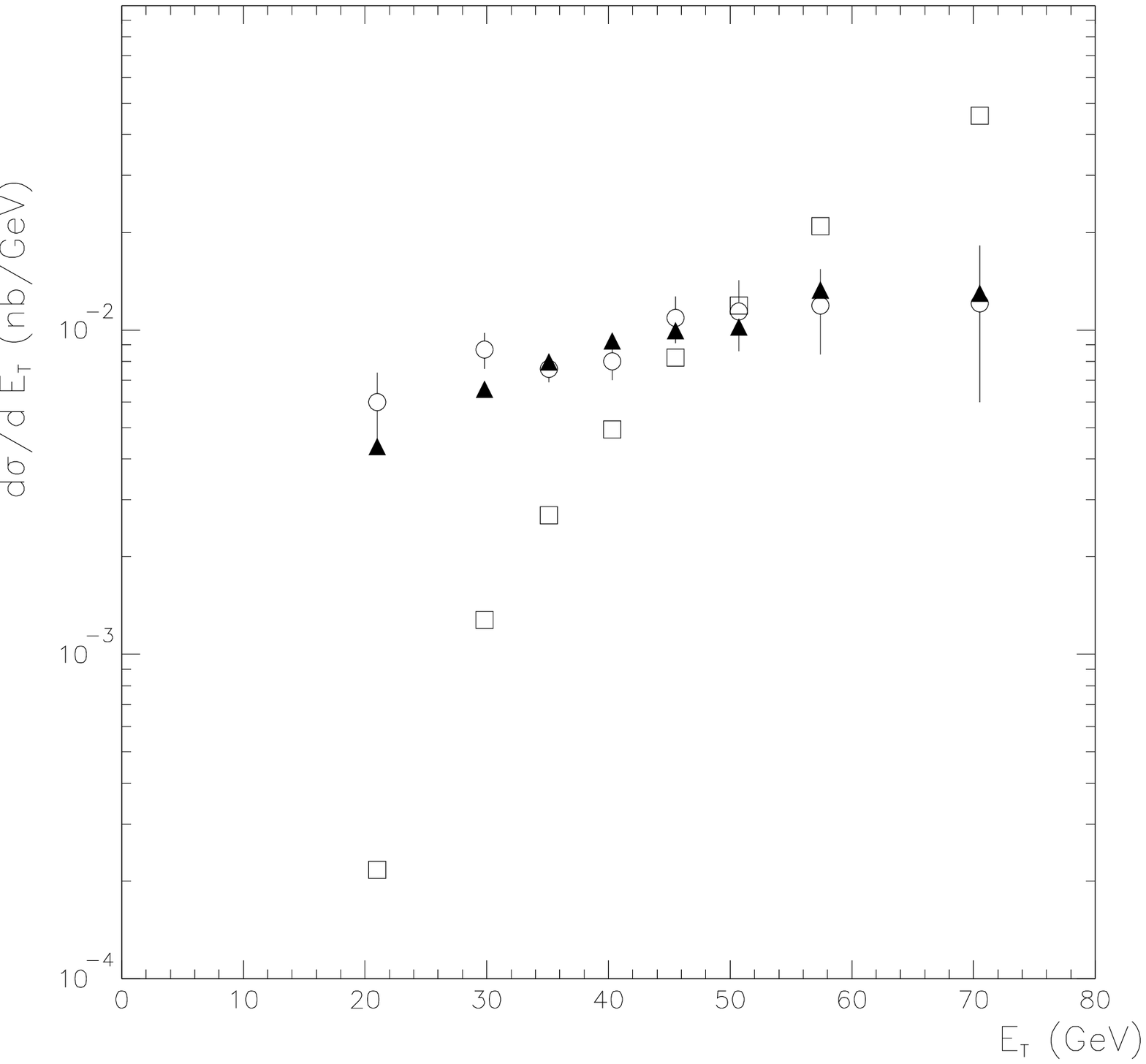,width=\linewidth}}
 \end{center}
 \caption{Comparison of the experimental $E_T$ spectrum of the rapidity gap
   fraction (open circles) with the $\gamma_B$ predictions for $m_b =$ 0 and
   70 GeV (solid triangle and open squares, respectively) .}
\label{fit:qu}
\end{figure}

In our analysis, we fixed the value of $m_B$ and determined $\alpha_B$
in order to fit the experimental $E_T$ spectrum, using the cuts and
$E_T$ bins defined in Ref.\ \cite{jill}. We exhibit in Fig.\
\ref{tev:gap} the region in the $(\alpha_B , m_B)$ plane obtained from
the fits to the data. Although the $\chi^2$ distribution as a function
of $\alpha_B$ has a well-defined minimum for all values of $m_B$, the
quality of the fit is good only at small $m_B$. This can be seen from
Fig.\ \ref{fit:qu}, which shows the general trend of the predicted
$E_T$ distribution as $m_B$ increases. Therefore, $\gamma_B$ exchange
cannot be the sole source of rapidity gap events at the Tevatron since
the low mass allowed region in Fig.\ \ref{tev:gap} has already been
ruled out by the direct $\gamma_B$ search in $\Upsilon$ decays
\cite{mu:3}. Furthermore, for larger $\gamma_B$ masses the fitted
values of $\alpha_B$ fall well within the region that can be probed
by the the direct search for $\gamma_B$ using Run I data, see Fig.\
\ref{fig:bb}. It is interesting to notice that introducing a gap
survival probability \cite{bj} $P_s < 1$ only worsens the problem
since this will require larger values of $\alpha_B$ to fit the data,
$\alpha_B \propto 1/\sqrt{P_s}$.

Hard jets separated by a rapidity gap have also been observed in
photoproduction events at the $ep$ collider HERA \cite{hera}. The
ZEUS Collaboration measured that approximately 10\% of the two-jet
events with $p_T > 6$ GeV and $\Delta \eta > 3$ present a rapidity gap
\cite{hera}. Assuming that these events stem from $\gamma_B$ exchange
in the $t$--channel we can constrain the mass and coupling of
$\gamma_B$. We present in Table \ref{gap:hera} the values of
$\alpha_B$ that lead to the observed gap fraction of 0.1 at large
rapidity separation, where we imposed the cuts of Ref.\ \cite{hera}.
Here it is also clear that these events cannot be explained solely as
being due to $\gamma_B$ exchange.

%******************************************************************************

\section{Conclusions}

In this work we demonstrated that the presently available Tevatron
data can be used to rule out the existence of bosons coupling to
baryon number for masses in the range 40--300 GeV and $\alpha_B
\gtrsim 0.02$, if no signal is found. In the near future, the Tevatron
experiments should increase this sensitivity to $\alpha_B \gtrsim
0.003$ at the Main Injector. These bounds would preclude $\gamma_B$
boson exchange as a significant source of events with rapidity gaps
between hard jets.

%******************************************************************************

\section{Acknowledgments}

We would like to thank T.\ Stelzer for helping us to introduce the
$\gamma_B$ into the package Madgraph. This work was supported by
Conselho Nacional de Desenvolvimento Cient\'{\i}fico e Tecnol\'ogico
(CNPq), by Funda\c{c}\~ao de Amparo \`a Pesquisa do Estado de S\~ao
Paulo (FAPESP), and by Programa de Apoio a N\'ucleos de Excel\^encia
(PRONEX).

%******************************************************************************

%%%%%%%%%%%%%%%%%%% Tables %%%%%%%%%%%%%%

\begin{table}
\begin{tabular}{||c|c||}
  $m_B$ (GeV)  &  $\alpha_B$ \\
\hline 
\hline
10.            &  1.05   \\
20.            &  2.84   \\
30.            &  4.59
\end{tabular}
\medskip\medskip
\caption{Values of $\alpha_B$ needed to explain the formation of
rapidity gaps in photoproduction events at HERA for several $\gamma_B$ masses.}
\label{gap:hera}
\end{table}

%%%%%%%%%%%%%%%%%%%% Figures %%%%%%%%%%%%%

%%%

%%%

%%%

%%%

%%%

%******************************************************************************

\begin{thebibliography}{99}
  
\bibitem{qg} S.\ Giddings and A.\ Strominger, Nucl.\ Phys.\ {\bf B307} (1988)
  854; S.\ Coleman, Nucl.\ Phys.\ {\bf B336} (1988) 643.
  
\bibitem{mu:1} H.\ Murayama and D.\ B.\ Kaplan, Phys.\ Lett.\ {\bf B336}
  (1994) 221; V.\ Ben-Hamo and Y.\ Nir, Phys.\ Lett.\ {\bf B339} (1994) 77;
  A.\ E.\ Faraggi, Nucl.\ Phys.\ {\bf B428} (1994) 111.
  
\bibitem{st:low} G.\ Shiu and S.-H.\ Henry Tye, preprint CNLS 98/1561
  (hep-th/9805157).
  
\bibitem{mu:2} For a model with a small mixing between $\gamma_B$ and
  electroweak bosons see C.\ D.\ Carone and H.\ Murayama, Phys.\ Rev.\ {\bf
    D52} (1995) 484.
  
\bibitem{mu:3}C.\ D.\ Carone and H.\ Murayama, Phys.\ Rev.\ Lett.\ {\bf 74}
  (1995) 3122.
  
\bibitem{cdf:bb} F.\ Abe {\em et al.}, CDF Collaboration, hep-ex/9809022.
  
\bibitem{jill} J.\ Perkins (D\O\ Collaboration), {\em Proceedings of the 5th
    International Workshop on Deep Inelastic Scattering and QCD}, Chicago,
  Illinois, 1997; FERMILAB-Conf-97/250-E. See also G.\ Snow, contribution to
  the {\em Proceedings of the International Conference on High Energy Physics
    1998}.
  
\bibitem{cdf:zbb} T.\ Dorigo for the CDF Collaboration, hep-ex/9806022.
  
\bibitem{madg} W.\ Long and T.\ Stelzer, Comput.\ Phys.\ Commun.\ {\bf 81}
  (1994) 357.
  
\bibitem{cdf-prl} S.\ Abachi {\em et al.} (D\O\ Collaboration), Phys.\ Rev.\ 
  Lett.\ {\bf 72} (1994) 2332; F.\ Abe {\it et al.}  (CDF Collaboration),
  Phys.\ Rev.\ Lett.\ {\bf 74} (1995) 855; Phys.\ Rev.\ Lett.\ {\bf 80} (1998)
  1156; K.\ Goulianos (CDF Collaboration), Proceedings of the LAFEX
  International School on High Energy Physics (LISHEP-98), Rio de Janeiro,
  Brazil, 1998, FERMILAB-CONF-98/118-E.
  
\bibitem{two} O.\ J.\ P.\ \'Eboli, E.\ M.\ Gregores, and F.\ Halzen, Phys.\ 
  Rev.\ {\bf D58} (1998) 114005.
  
\bibitem{bj} J.\ D.\ Bjorken, Int.\ J.\ Mod.\ Phys.\ {\bf A7} (1992) 4189;
  Phys.\ Rev.\ {\bf D47} (1993) 101; preprint SLAC-PUB-5823 (1992).
  
\bibitem{hera}M.\ Derrick {\em et al.}, ZEUS Collaboration, Phys.\ Lett.\ {\bf
    B369} (1996) 55; T.\ Ahmed, {\em et al.}, H1 Collaboration, Nucl.\ Phys.\ 
  {\bf B435} (1995) 3.


\end{thebibliography}
\end{document}